\documentclass[conference]{IEEEtran}
\IEEEoverridecommandlockouts

\usepackage{cite}
\usepackage{amsmath,amssymb,amsfonts,thmtools}
\usepackage[noend]{algorithmic}
\usepackage[pdftex]{graphicx}
\usepackage{xcolor}
\def\BibTeX{{\rm B\kern-.05em{\sc i\kern-.025em b}\kern-.08em
    T\kern-.1667em\lower.7ex\hbox{E}\kern-.125emX}}

\usepackage[hyphens]{url}

\usepackage{algorithm}
\usepackage[caption=false]{subfig}

\usepackage{enumerate}

\usepackage{color}
\definecolor{red}{rgb}{1,0.2,0.2}
\definecolor{green}{rgb}{0.2,1,0.5}
\definecolor{blue}{rgb}{0,0,1}
\definecolor{lightblue}{rgb}{0.3,0.5,1}

\newtheorem{theorem}{Theorem}

\begin{document}

\title{Online Service Caching and Routing at the Edge with Unknown Arrivals}

\author{\IEEEauthorblockN{Siqi Fan, I-Hong Hou}
\IEEEauthorblockA{
\textit{Texas A\&M University}\\
College Station, USA \\
\{siqifan, ihou\}@tamu.edu}
\and
\IEEEauthorblockN{Van Sy Mai, Lotfi Benmohamed}
\IEEEauthorblockA{\textit{National Institute of Standards and Technology} \\
Gaithersburg, USA \\
\{vansy.mai, lotfi.benmohamed\}@nist.gov}
}

\maketitle

\begin{abstract}
  This paper studies a problem of jointly optimizing two important operations in mobile edge computing without knowing future requests, namely service caching, which determines which services to be hosted at the edge, and service routing, which determines which requests to be processed locally at the edge. We aim to address several practical challenges, including limited storage and computation capacities of edge servers and unknown future request arrival patterns. To this end, we formulate the problem as an online optimization problem, in which the objective function includes costs of forwarding requests, processing requests, and reconfiguring edge servers. By leveraging a natural timescale separation between service routing and service caching, namely, the former happens faster than the latter, we propose an online two-stage algorithm and its randomized variant. Both algorithms have low complexity, and our fractional solution achieves sublinear regret. Simulation results show that our algorithms significantly outperform other state-of-the-art online policies. 
\end{abstract}



\section{Introduction}
\label{section:introduction}

A growing challenge for mobile computing is the proliferation of data/computation-intensive and delay-sensitive applications, such as cognitive assistance and augmented reality (AR). On the one hand, running these applications completely within mobile devices may be infeasible due to the limited computation, storage, and battery capacity of such devices. On the other hand, offloading computation tasks of these applications to remote data centers may result in excessive end-to-end latency and hence poor user experience.

Such a dilemma has given rise to the popularity of mobile edge computing \cite{Kitanov2016fog, beck2014mobile}. In mobile edge computing, edge servers are deployed close to wireless base stations. These servers can host some popular services and process the corresponding computation tasks directly without having to forward them to remote data centers. Due to their close proximity to end users, edge servers are able to provide these services with much lower latency.

Despite the obvious advantage of mobile edge computing, there remain multiple important challenges 
that need to be addressed. 
First, edge servers can often host (or cache) a small number of services, and installing new services is typically time-consuming and expensive since it involves downloading all necessary data from remote data centers and setting up appropriate virtual machines or containers.
Second, edge servers usually have limited computation power, and hence requests will suffer queuing delays. So, the edge server needs to decide whether to process a request locally or not, even it has already cached the corresponding service.
Third, mobile users generate requests for services in arbitrary and typically time-varying patterns, which are hard to learn. Thus, edge servers must decide which services to cache and which requests to process without knowledge of future requests.

Most existing works only focus on one or two challenges above. Some studies only address the online caching problem with unknown request arrival patterns. For example, Paschos \textit{et al.} \cite{paschos2019learning, paschos2020online}, and Gao \textit{et al.} \cite{gao2020proactive} propose online algorithms with sublinear regret based on the online gradient ascent and bandit learning method. Zhao \textit{et al.} \cite{Zhao2018red} address the installation cost and analyze the competitive ratio of their algorithm. 
These studies fail to take the limited computation power of edge servers into account. Some other papers consider joint designs of service caching and request processing by explicitly address limits on both storage and computation power. For instance, Li \textit{et al.} \cite{li2021cooperative} and Xu \textit{et al.} \cite{jie2018caching} propose online algorithms for optimizing service caching and request routing based on a Lyapunov optimization framework. A weakness of these caching and routing solutions is that they assume that the request arrival patterns are predictable or follow a certain stationary random process.

In this paper, we aim to address all three challenges and minimize a combination of the queuing latency, forwarding latency, and installation costs. By leveraging the natural timescale separation between service caching and service routing, we formulate the problem into a two-stage online optimization problem without knowledge of future requests. 

To solve this problem, we propose a low-complexity two-stage online policy and its randomized variant. The two-stage online policy consists of two parts: the first part is a low-complexity algorithm that finds not only the optimal service routing decision but also the gradient of the current service caching decision, despite that neither of them have closed-form expressions, and the second part employs online projected gradient descent to update the service caching decisions.  
We further design its randomized variant to make the probabilistic caching solution from the two-stage online policy implementable. We theoretically prove that our two-stage online policy achieves sublinear regret, while the randomized one at most triples the installation cost.


Our online algorithms are evaluated through simulations under various scenarios. We compare them against three other algorithms, including an offline policy that knows overall request popularity in advance. Results show that our algorithms perform much better than other online algorithms and perform virtually the same as the offline policy.

The rest of the paper is organized as follows.  Section \ref{section:model} introduces our system model and problem formulation. Section \ref{sec:algorithm} presents our two-stage online algorithm for obtaining fractional solutions with sublinear regret. Section \ref{section:integral} includes a randomized variant of the algorithm that ensures integer solutions for the service caching problem. Section \ref{section:simulation} shows our simulation results under a variety of scenarios. Finally, Section \ref{section:conclusion} concludes the paper.


\section{System Model}
\label{section:model}

\subsection{System Overview}

We consider an edge system with a backhaul connection. This edge system includes multiple clients, an edge server, and remote data centers. Clients generate requests for different services according to some unknown and unpredictable patterns, and then send these requests to the edge server. 
The edge server may \emph{cache} some services and process some requests for these services locally, while forwarding remaining requests to remote data centers. Requests processed at the edge encounter a \emph{computation latency} due to the limited computation capacity of the edge server, while requests forwarded to remote data centers encounter a \emph{forwarding latency} due to network latency.

\subsection{Service Caching and Processing}

We assume that time is slotted, and the system runs for $T$ time slots. The duration of a time slot is chosen so that, in any given time slot, the patterns for the service requests (originating from different clients) remain roughly the same.

We use $N$ to denote the total number of different services. 
Let $x_{n,t}\in\{1,0\}$ be a binary decision variable that indicates whether the edge server caches service $n$ in time slot $t$, and let $X_{t}:=[x_{1,t},x_{2,t},\dots,x_{N,t}]$. Since the edge server has limited storage capacity, we assume that the edge server can cache at most $Z$ services, i.e., $\sum_{n=1}^{N} x_{n,t}\leq Z,\forall t$. We call the problem of determining $X_t$ the \emph{service caching problem}. 

Caching a new service at the edge can be a costly process, which typically involves downloading codes and databases and setting up virtual machines or containers. Thus, we assume that the edge server must decide $X_t$ before time slot $t$.

At the beginning of time slot $t$, the edge server observes the requests from clients and calculates the request arrival rates. We use $\lambda_{n,t}$ to denote the number of requests for service $n$ in time slot $t$, and let $\Lambda_t:=[\lambda_{1,t},\lambda_{2,t},\dots,\lambda_{N,t}]$. We assume that an upper bound $W$ on the total arrival rate is known, that is, $\sum_{n=1}^{N}\lambda_{n,t}\leq W,\forall t$.


During each time slot $t$, the edge server needs to decide which requests to be processed locally. Due to the limited computation power of the edge server, it may not be desirable to process all requests for services that it caches. For a service $n$, the edge server will process a fraction $y_{n,t} \in [0,1]$ of the number of requests locally, and forward the remaining $(1-y_{n,t})$ portion of the requests to the data center. Since the edge server can only process requests whose corresponding services have already been cached at the edge, we require that $y_{n,t}\leq x_{n,t}, \forall n$.

Let $Y_t:=[y_{1,t},y_{2,t},\dots,y_{N,t}]$.  
We call the problem of determining $Y_t$ the \emph{service routing problem}. Since the edge server can adjust service routing in real time, we consider that the edge server determines $Y_t$ after it observes $\Lambda_t$.

\subsection{Cost and Problem Formulation}

The goal of the edge server is to minimize the total cost of the system, which consists of \emph{latency cost} and \emph{installation cost}, by jointly optimizing $X_t$ and $Y_t$.

First, the latency cost refers to the total latency experienced by all requests. In the system, when a request is forwarded to the remote data center, it experiences a forwarding latency, which is denoted as $d_{n}$ for service $n$. When a request is processed at the edge, it experiences a computation latency due to the limited computation power of the edge server. It is reasonable to assume that the per-request computation latency at the edge depends on the total computation load, and can be described by a convex, increasing, and differentiable function $C(\cdot)$ with $0 \le C(0) \leq d_n, \forall n$ and $\lim_{s\to\infty} C(s) = \infty$. Since the total computation load at the edge server is $\sum_{n=1}^{N}\lambda_{n,t}y_{n,t}$ in time slot $t$, the total latency of all requests can be written as
\begin{align*}
    L_t(Y_t) :=\!\! \sum_{n=1}^{N} \!\lambda_{n,t}y_{n,t} C\big(\! \sum_{m=1}^{N}\lambda_{m,t}y_{m,t}\big) \!+\!\! \sum_{n=1}^{N} \!\lambda_{n,t}(1 \!-\! y_{n,t})d_n.
\end{align*}
The goal of the edge server at each time slot $t$ is to minimize this latency cost given the current caching decision $X_t$, i.e., to solve
\begin{align}
    G_t(X_t):=\min_{Y_t} \quad& L_t(Y_t), \label{eq:rout_obj}\\
    \textrm{s.t.} \quad& 0\le y_{n,t}\leq x_{n,t}, \quad \forall n. \label{eq:rout_cons1}
\end{align}

\begin{figure}[thb]
    \centering
    \includegraphics[scale=0.175]{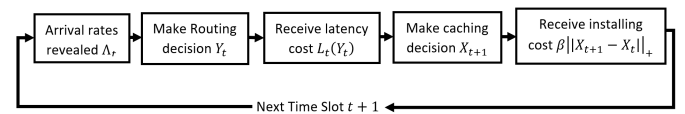}
    \caption{Process of online service caching and routing.}
    \label{fig:processes}
\end{figure}

Second, the installation cost refers to the operation cost incurred when the edge server caches new services. 
For simplicity, 
we assume that caching every new service incurs a cost of $\beta$. Hence, the total installation cost is $\beta \sum_{t=1}^{T}\|X_t-X_{t-1}\|_+$, where $\|X_t\|_+=\sum_{n=1}^{N}\max\{x_{n,t},0\}$.
As a result, the sequence of determining variables and receiving cost is illustrated in  Fig.~\ref{fig:processes}, and the total cost over $T$ time slots can be written as $\sum_{t=1}^{T}(G_t(X_t)+\beta \|X_t-X_{t-1}\|_+)$.

Here, it is important to note that the service caching problem and the service routing problem operate on different timescales. The edge server needs to decide $X_t$ in time slot $t-1$, without any knowledge about $\Lambda_t$. In contrast, the edge server can decide the value of $Y_t$ after observing the first few requests in time slot $t$ to estimate the arrival rate $\Lambda_t$.

The edge server aims to find $[X_1,X_2,\dots,X_T]$ and $[Y_1,Y_2,\dots,Y_T]$ that minimize the total cost subject to all the constraints described above. 
However, since this is a mixed integer problem due to binary caching decisions $x_{n,t}\in\{0,1\}$, we first relax this constraint and allow $x_{n,t}$ to be any real number in $[0,1]$. Under this relaxation, $x_{n,t}$ can be interpreted as the probability of caching service $n$ at time $t$, and the (offline) problem of minimizing the total cost becomes:
\begin{align}
    \min_{[X_t]} \quad & \sum_{t=1}^{T}\Big(G_t(X_t)+\beta \|X_t-X_{t-1}\|_+ \Big), \label{equation:OCO_intro} \\
    \textrm{s.t.} \quad& 0\leq x_{n,t}\leq 1, \quad \forall n, \forall t, \\
    & \sum_{n=1}^{N}x_{n,t}\leq Z, \quad  \forall t.
\end{align}

While the above problem is a convex optimization problem, solving it requires the knowledge of all request arrival rates, that is, all $\Lambda_t$, in advance. In practice, however, the edge server needs to make service caching  decision $X_t$ without the knowledge of future arrival rates, hence the need for an \emph{online algorithm}.


The performance of an online algorithm that does not have any knowledge about future arrivals is evaluated by comparing it against an offline policy that knows future arrivals. Let us consider an offline policy that knows the overall popularity of all services, i.e., $\sum_t \lambda_{n,t}$ for all $n\in [1,N]$, but not individual $\lambda_{n,t}$. In this case, the optimal static caching policy will cache $Z$ services with largest $d_n\sum_t \lambda_{n,t}$ at all times. The assumption for this offline optimal static policy is widely used in caching literature; see, e.g., \cite{bhattacharjee2020fundamental}. Let $X^o:=[x^{o}_1, x^{o}_2, \dots, x^{o}_N]$ be the service caching decision of this offline policy. Note that, since we assume that the edge server can adjust service routing in real time, the offline policy can determine an optimal routing decision $[Y_t^o]$ that minimizes the latency cost $L_t(Y_t)$ in each time slot $t$ while satisfying the constraint $y^o_{n,t}\leq x_n^o,\forall n$.

Let $\hat{X}:=[\hat{X}_1,\hat{X}_2,\dots,\hat{X}_T]$ and $\hat{Y}:=[\hat{Y}_1,\hat{Y}_2,\dots,\hat{Y}_T]$ be solutions produced by an online algorithm $\xi$, then we define the regret of $\xi$ as the difference between its cost and the cost of the  optimal static offline policy:
\begin{equation*}
    Reg(\xi) := \sum_{t=1}^{T}\big(L_t(\hat{Y}_t) - L_t(Y_t^{o}) +\beta\|\hat{X}_t-\hat{X}_{t-1}\|_+\big).
\end{equation*}
It should be noted that the offline policy doesn't incur installation costs since it uses a fixed caching decision.

The goal of this paper is to find an online algorithm with provably small regret under any sequence of arrival rates.


\section{Online Algorithm}
\label{sec:algorithm}

In this section, we present a two-stage online algorithm that aims to jointly optimize the service caching and routing decisions asymptotically. After arrival rates are revealed in each time slot, the first stage computes routing decisions and produces necessary parameters for the next stage, which updates caching decisions for the next time slot.

\subsection{Optimal Routing}

In this subsection, we consider the service routing problem at time $t$ described in \eqref{eq:rout_obj}--\eqref{eq:rout_cons1}, given the current caching decision $X_t$ and arrival rates $\Lambda_t$.  
Solving this problem with a general convex optimization solver may, however, incur high complexity. Surprisingly, we show below that there exists an $O(N)$ algorithm that not only solves this problem but also provides a subgradient $\nabla G_t(X_t)$, which is important for dealing with the service caching problem in the second stage.

The main idea of our algorithm is to leverage the special structure in $L_t(Y_t)$. Let 
\begin{align*}
    J_t(Y_t) := C(\sum_{n=1}^{N}\lambda_{n,t}y_{n,t})+\sum_{n=1}^{N}\lambda_{n,t}y_{n,t} C^{'}(\sum_{m=1}^{N}\lambda_{m,t}y_{m,t}).
\end{align*}
Then, we have $\frac{1}{\lambda_{n,t}}\frac{\partial L_t(Y_t)}{\partial y_{n,t}} = J_t(Y_t) -d_n$, which corresponds to the marginal benefit of processing one more request for service $n$ at the edge. Sorting all services so that $d_1\geq d_2\geq\dots\geq d_N$, then we have $\frac{1}{\lambda_{1,t}}\frac{\partial L_t(Y_t)}{\partial y_{1,t}}\leq \frac{1}{\lambda_{2,t}}\frac{\partial L_t(Y_t)}{\partial y_{2,t}}\leq\dots\leq\frac{1}{\lambda_{N,t}}\frac{\partial L_t(Y_t)}{\partial y_{N,t}}$. Based on this observation, we design Algorithm~\ref{alg:routing} shown below.

\begin{algorithm}   
\caption{\texttt{ServiceRouting}}
\begin{algorithmic}[1]
\renewcommand{\algorithmicrequire}{\textbf{Input:}}
\REQUIRE $d_1 \ge  d_2 \ge \ldots \ge d_N$, $X_t, \Lambda_t$
\renewcommand{\algorithmicrequire}{\textbf{Initialize:}}
\REQUIRE $Y_t\leftarrow 0$
\FOR{$n=1,2,...,N$}
\IF{$J_t(Y_t)-d_n<0$}
\STATE $y_{n,t}\leftarrow x_{n,t}$
\IF{$J_t(Y_t)-d_n>0$}
\STATE choose $y_{n,t} \in [0,x_{n,t}]$ s.t. $J_t(Y_t)-d_n=0$
\ENDIF
\ENDIF
\ENDFOR
\FOR{$n=1,2,...,N$}
\IF{$J_t(Y_t) \leq d_n$}
\STATE $\nu_n\leftarrow \lambda_{n,t}(d_n-J_t(Y_t))$, $\mu_n\leftarrow 0$
\ELSE
\STATE $\nu_n\leftarrow 0$, $\mu_n\leftarrow \lambda_{n,t}(J_t(Y_t)-d_n)$
\ENDIF
\ENDFOR
\renewcommand{\algorithmicrequire}{\textbf{Output:}}
\REQUIRE $Y_t, \nabla G_t(X_t) \leftarrow [-\nu_1, -\nu_2,\dots,-\nu_N]$
\end{algorithmic}
\label{alg:routing}
\end{algorithm}

\begin{theorem}\label{thmRouting}
    Algorithm~\ref{alg:routing} produces an optimal solution for the routing problem \eqref{eq:rout_obj}--\eqref{eq:rout_cons1} and a subgradient $\nabla G_t(X_t)$ is given by
    \begin{equation}
        \frac{\partial G_t(X_t)}{\partial x_{n,t}}  = -  \nu_n  = 
        \begin{cases}
            &  \lambda_{n,t}(J_t(Y_t)-d_n), \textbf{ if } y_{n,t} = x_{n,t}, \\
            & 0, \qquad\qquad\qquad\quad\; \textbf{otherwise}.
        \end{cases}
        \nonumber
    \end{equation}
\end{theorem}
\begin{IEEEproof}
    First, it can be verified that $Y_t, \nu_n$ and $\mu_n$ produced by Algorithm~\ref{alg:routing} satisfy the KKT conditions of problem  \eqref{eq:rout_obj}--\eqref{eq:rout_cons1}, i.e.,
    \begin{align}
        \lambda_{n,t}(J_t(Y_t) - d_n) - \mu_n + \nu_n = 0, \forall n, \label{eq:KKT1} \\
        \nu_n (y_{n,t} - x_{n,t})=0, \mu_{n}(-y_{n,t}) = 0, \forall n, \label{eq:KKT2}\\
        \mu_n \geq 0,\; \nu_n \geq 0, \forall n, \label{eq:KKT3}
    \end{align}
    where $\nu_n$ and $\mu_n$ are Lagrange multipliers associated with the constraints in \eqref{eq:rout_cons1}. Since the problem is convex, it follows that $Y_t$ is an optimal solution.

    Second, it follows from \cite[\S 5.6]{Boyd2004Conv} that $G_t$ is convex in $X_t$ and that $\frac{\partial G_t(X_t)}{\partial x_{n,t}}  = -  \nu_n,\forall n=1,\ldots,N$. 
    This completes the proof. 
\end{IEEEproof}


\subsection{Online Service Caching}
For service caching, we adopt the online gradient descent method with lazy projection in \cite{Shai2012OCO}, where the update step at time $t$ is given in Algorithm~\ref{alg:caching} below. Here, $\nabla G_t(X_t)$ is the subgradient calculated by Algorithm~\ref{alg:routing}, $\eta$ is the step size, and $\theta_t=[\theta_{1,t},\theta_{2,t},\dots,\theta_{N,t}]$ is an internal vector with $\theta_1=\textbf{0}$.


\vspace{-0.5em}
\begin{algorithm}   
\caption{\texttt{ServiceCaching}}
\begin{algorithmic}[1]
\renewcommand{\algorithmicrequire}{\textbf{Input:}}
\REQUIRE $\theta_{t},\nabla G_{t}(X_{t}),\eta$
\STATE $\theta_{t+1} \gets \theta_{t}-\nabla G_{t}(X_{t})$
\STATE $X_{t+1} \leftarrow$ the Euclidean projection of $\eta\theta_{t+1}$ onto the set $\{X\in \mathbb{R}^N~|~0\leq x_{n}\leq 1, \; \sum_{n=1}^{N}x_{n}\leq Z\}$
\renewcommand{\algorithmicrequire}{\textbf{Output:}}
\REQUIRE $X_{t+1},\theta_{t+1}$
\end{algorithmic}
\label{alg:caching}
\end{algorithm}
\vspace{-1em}


As a result, combining both Algorithms~\ref{alg:routing} and \ref{alg:caching} yields an online service caching and routing algorithm (OCR) shown in  Algorithm~\ref{alg:OCR} below. 

\vspace{-0.5em}
\begin{algorithm}
\caption{Online service Caching and Routing (OCR)}
\begin{algorithmic}[1]
\renewcommand{\algorithmicrequire}{\textbf{Initialize:}}
\REQUIRE $\eta$, $\theta_1\leftarrow 0$, $X_1$
\FOR{$t=1,2,...,T$}
\STATE $Y_t, \nabla G_t(X_t)\leftarrow \texttt{ServiceRouting}(X_t,\Lambda_t)$
\STATE $X_{t+1},\theta_{t+1} \leftarrow \texttt{ServiceCaching}(\theta_{t},\nabla G_{t}(X_{t}),\eta)$
\ENDFOR
\end{algorithmic}
\label{alg:OCR}
\end{algorithm}

\vspace{-1em}

Next we show that OCR achieves a sublinear regret. 
\begin{theorem}
    Let $\eta=O(\frac{1}{\sqrt{T}})$. 
    Then, $Reg(OCR)=O(\sqrt{T})$.
\end{theorem}
\begin{IEEEproof}
Since Algorithm~3 can be viewed as applying online gradient descent with lazy projection to the problem in \eqref{equation:OCO_intro} with objective function $G_t(X_t)$ (i.e., without the installation cost), it follows from  \cite[Corollary~2.17]{Shai2012OCO} that  the regret (in terms of $G_t(X_t)$) is $O(\sqrt{T})$ when $\eta=O(\frac{1}{\sqrt{T}})$, provided that $\nabla G_t$ is bounded. In our case, 
the boundedness holds because
\begin{equation}
\|\nabla G_t(X_t)\|_2^2 = \sum_{n=1}^{N} \nu_n^2 \leq \sum_{n=1}^{N} \lambda_{n,t}^2 d_n^2 \leq W^2 \max_i d_i^2, \label{eqGradBound}
\end{equation}
where we have used Theorem~\ref{thmRouting} and the fact that $\sum_{n=1}^{N} \lambda_{n,t}^2\le \big(\sum_{n=1}^{N} \lambda_{n,t}\big)^2 = W^2$. 

    It remains to show that $\beta \sum_{t=1}^T \|X_t-X_{t-1}\|_+=O(\sqrt{T})$. To this end, 
    note that 
    $\|X_t-X_{t-1}\|_+ 
    \leq  \|X_t-X_{t-1}\|_1 
    \leq \sqrt{N}\|X_t-X_{t-1}\|_2 
    \leq \sqrt{N}\|\eta \nabla G_t(X_t)\|_{2}$,
    where the last inequality follows from Algorithm~\ref{alg:caching} and the nonexpansiveness property of Euclidean projections. 
    Next, using \eqref{eqGradBound} and the fact that $\eta=O(\frac{1}{\sqrt{T}})$, we have $\sum_{t=1}^{T}\beta \|X_t-X_{t-1}\|_+ \leq \sum_{t=1}^{T} \eta\beta\sqrt{N}\| \nabla G_t(X_t)\|_{2} \leq T \eta \beta \sqrt{N} \max_i d_i = O(\sqrt{T}).$
    
    Thus, we conclude that $Reg(OCR)= O(\sqrt{T})$.
\end{IEEEproof}

    
    

Finally, we analyze the complexity of Algorithm~\ref{alg:OCR}. It can be seen that the bottleneck is the projection step in the line 2 of Algorithm~\ref{alg:caching}.

We consider the following steps for finding the projection of $\eta \theta_{n,t+1}$ onto the set $\{X\in \mathbb{R}^N~|~0\leq x_{n}\leq 1, \; \sum_{n=1}^{N}x_{n}\leq Z\}$. Let $X'$ be the vector of $x'_n$ where $x'_n = \min\{1,\max\{0,\eta \theta_{n,t+1}\}\}$. If $\sum_{n=1}^N x'_n\leq Z$, then $X'$ is the projection. Otherwise, the projection, denoted by $X^*$, must have $\sum_{n=1}^N x^*_n=Z$. Then, we can employ the algorithm in \cite{wang2015projection}, which has complexity $O(N^2)$, to obtain $X^*$. 
Hence, the overall complexity of Algorithm 3 is $O(N^2)$ per time slot.



\section{Randomized Algorithm for Service Caching}
\label{section:integral}

The online algorithm for finding $X_t$ as proposed in Algorithm~\ref{alg:caching} may produce fractional solutions, which can be interpreted as the probability that the edge server caches each service. In this section, we propose a randomized algorithm that satisfies this probability interpretation while guaranteeing a provably small installation cost.

\subsection{Randomized Algorithm}

The basic idea of our randomized algorithm is to simultaneously maintain $K$ sample paths, where each sample path represents a probability mass of $\frac{1}{K}$. We then quantize each $x_{n,t}$ into a multiple of $\frac{1}{K}$. Specifically, let $X^Q_t$ be the quantized version of $X_t$, we then require that $Kx^Q_{n,t}$ to be a non-negative integer and $\sum_n x^Q_{n,t}\leq Z$.

Let $r_{k,n,t}$ be the indicator function that service $n$ is cached at the edge at time $t$ in the sample path $k$. Let $R_{k,t}$ be the vector $[r_{k,1,t}, r_{k,2, t}, \dots]$. In every time slot $t$, our randomized algorithm receives $X^Q_t$ from Algorithm~\ref{alg:caching}. Then, it constructs $R_{k,t}$ based on $X^Q_t$ and $R_{k,t-1}$ to ensure three properties: First, the probability of caching service $n$ is indeed $x^Q_{n,t}$, that is, $\sum_{k=1}^K r_{k,n,t} = Kx^Q_{n,t}$. Second, the storage capacity constraint is satisfied for all sample paths, that is, $\sum_n r_{k,n,t}\leq Z, \forall k$. Third, the expected installation cost, which can be expressed as $\frac{1}{K}\sum_k \|R_{k,t}-R_{k,t-1}\|_+$, is bounded. Let $\Delta_t:=[\delta_{1,t},\delta_{2,t},\dots,\delta_{N,t}]$ be the difference between $X_t^Q$ and $X_{t-1}^Q$. Algorithm~\ref{alg:rand} shows the complete randomized algorithm, including decisions on service caching and routing.

\begin{algorithm} 
\caption{Randomized Online service Caching and Routing (ROCR)} \label{alg:rand}
\begin{algorithmic}[1]
\renewcommand{\algorithmicrequire}{\textbf{Initialize:}}
\REQUIRE $K$, $R_{k,1}\leftarrow 0, \forall k, \eta, \theta_1 \leftarrow 0$
\STATE Choose $k^*$ uniformly from $\{1,2,\dots,K\}$.
\FOR{$t=1,2,...,T$}
\STATE Observe $\Lambda_t$.
\STATE $Y_t,\nabla G_t(X_t) \leftarrow \texttt{ServiceRouting}(R_{k^*,t},\Lambda_t)$.
\STATE $X_{t+1}^Q, \theta_{t+1} \leftarrow \texttt{ServiceCaching}( \theta_{t}, $ $ \nabla G_{t}(X_{t}),  \eta)$.
\STATE $R_{k,t+1} \leftarrow R_{k,t}, \forall k$.
\STATE $\Delta_{t+1} \leftarrow$ $X_{t+1}^Q-X_{t}^Q$.
\FOR{$n=1,2,\dots,N$}
\IF{$\delta_{n,t+1}>0$}
\STATE Randomly choose $K\delta_{n,t+1}$ sample paths with $r_{k,n,t+1}=0$, and set $r_{k,n,t+1}=1$ for them.
\ELSIF{$\delta_{n,t}<0$}
\STATE Randomly choose $|K\delta_{n,t+1}|$ sample paths with $r_{k,n,t+1}=1$, and set $r_{k,n,t+1}=0$ for them.
\ENDIF
\ENDFOR
\WHILE{$\exists \hat{k}$ such that $\sum_n r_{\hat{k},n,t+1}>Z$}
\STATE Find one sample path $k'$ with $\sum_n r_{k',n,t+1}<Z$.
\STATE Find a service $\hat{n}$ with $r_{\hat{k}, \hat{n},t+1}=1$, $r_{k', \hat{n},t+1}=0$.
\STATE Set $r_{\hat{k}, \hat{n},t+1}=0$ and $r_{k', \hat{n},t+1}=1$.
\ENDWHILE
\STATE Cache all services with $r_{k^*,n,t+1}=1$.
\ENDFOR
\end{algorithmic}
\label{algorithm:integral}
\end{algorithm}

By the design of Algorithm~\ref{alg:rand}, we obviously have the first two properties. We show below that Algorithm~\ref{alg:rand} also enjoys a provably small expected installation cost.

\subsection{Performance Analysis}

First, we consider the influence of Algorithm~\ref{alg:rand} on the installation cost, which is shown below.

\begin{theorem}
    The expected installation cost at each time slot in Algorithm~\ref{algorithm:integral} is at most $3\beta\|X_t^Q-X_{t-1}^Q\|_+$. 
    \label{theorem:integra_bound}
\end{theorem}

\begin{IEEEproof}
    As the installation cost only happens when we increase $r_{k,n,t}$, we aim to bound the increase in $r_{k,n,t}$. Under Algorithm~\ref{algorithm:integral}, $r_{k,n,t}$ can be changed either in lines 9--12 or in lines 14--16. In lines 9--12, the total increase is $K\|X_t^Q-X_{t-1}^Q\|_+$. Moreover, every change in lines 9--12 can result in at most two changes in lines 14--16. Hence, the total increase in lines 14--16 is at most $2K\|X_t^Q-X_{t-1}^Q\|_+$.
    
    Thus, the maximum increase in Algorithm~\ref{algorithm:integral} is $3K\|X_t^Q-X_{t-1}^Q\|_+$ over all sample paths. Since each sample path represents a probability mass of $\frac{1}{K}$, the expected installation cost is at most $3\beta\|X_t^Q-X_{t-1}^Q\|_+$.
\end{IEEEproof}

Next, we analyze the complexity of Algorithm~\ref{alg:rand}. Since $\sum_nX_t^Q \leq Z$ and  $\sum_nX_{t-1}^Q\leq Z$, at most $KZ$ variables will be increased to 1 and at most $KZ$ variables will be decreased to 0 in Steps 10--12. This is a total of $O(KZ)$ changes. To implement the while loop in Steps 14--16, we can first divide all sample paths into three groups: those with $\sum_n r_{\hat{k},n,t}>Z$, those with $\sum_n r_{\hat{k},n,t}=Z$, and those with $\sum_n r_{\hat{k},n,t}<Z$. Then, Step 14 is an $O(1)$ operation. Step 16 takes $O(N)$ time. We note that each increase in Steps 10--12 will result in at most one iteration of the while loop in Steps 14--16. Hence, steps 14--16 will be executed at most $KZ$ times and the overall complexity of this while loop is $O(KZN)$. 
Thus, the complexity of Algorithm~\ref{alg:rand} is $O(\max\{KZN,N^2\})$ per time slot.


\section{Simulation Results}
\label{section:simulation}

In this section, we conduct various simulations to evaluate the performance of our algorithms OCR 
and ROCR. 

\subsection{Setup}

We conduct experiments on following two datasets:
\begin{itemize}
    \item The first dataset is a synthetic dataset, following a random replacement model in 
    \cite{elayoubi2015performance,paschos2019learning}
    with $N=10^3$ and $T=10^4$. In this dataset, all requests follow a Zipf distribution, while the ranking of services frequently changes according to Table 2 in \cite{elayoubi2015performance}.
    \item The second dataset is based on the Google trace data from \cite{Google2010}, containing a sequence of different service requests. This dataset includes more than three million requests for $N=9,218$ services within a seven-hour timespan. As time is slotted in the trace data by 300 seconds, which is a large jump, we divide each interval into 300 different parts with an equal number of requests following the original sequence. In this dataset, the popularity of requests in one time slot changes fast, while some services are very popular over the whole time period.
\end{itemize}

Considering the queuing delay, we assume that the edge server operates like a $M/M/1$ queuing system \cite{Thomas1976Queue} with service rate $\phi$, i.e.,  $C(\sum_{i=1}^{N}y_{i,t})=\frac{1}{\phi-\sum_{i=1}^{N}y_{i,t}}$.

The system parameters are shown in Table~\ref{tab1}, where the values of  forwarding latency, service rate, and cache limit follow the parameters of services and base stations in \cite{jie2018caching}.

\vspace{-1em}
\begin{table}[htbp]
\caption{System Parameters}
\begin{center}
\begin{tabular}{|c|c|c|c|c|c|}
\hline
Parameter & \!\textbf{\textit{$d_n$}} (sec/request)\!& \!\textbf{\textit{$\phi$}} (request/sec) \!\!& \textbf{\textit{$Z$}} & \textbf{\textit{$K$}} & \textbf{\textit{$\eta$}} \\
\hline
Value & $[2,4]$  & $[20,100]$ & $[2,10]$ & $10^2$ & $0.05$\\
\hline
\end{tabular}
\label{tab1}
\end{center}
\end{table}
\vspace{-1em}


Throughout the evaluation, we compare ROCR and OCR with
the following baseline approaches:
\begin{itemize}
    \item OGA (Online Gradient Ascent \cite{paschos2019learning}):
    In each time slot, it uses $[\lambda_{1,t}d_1, \lambda_{2,t}d_2,\dots, \lambda_{N,t}d_N]$ as the gradient for the service caching problem. Since OGA does not consider the routing procedure, we apply our routing policy in this algorithm to obtain its best performance. OGA produces fractional $X_t$ and its cost is based on this fractional $X_t$.
    
    \item OFF (Offline Policy): This is the optimal static offline policy defined in Sec. II. It caches the same $Z$ services with the largest $\sum_{t=1}^{T}\lambda_{n,t}d_n$ in all time slots and applies optimal routing decisions.

    \item OREO: (Online se{R}vice caching for mobile {E}dge c{O}mputing \cite{jie2018caching}): This algorithm jointly optimizes service caching and routing decisions with energy and cost constraints. In the context of this work, all energy and cost constraints in \cite{jie2018caching} are relaxed to be infinite. As suggested by \cite{jie2018caching}, we let the arrivals of the current time slot be the prediction for the next time slot and use the Gibbs sampling method with parameter $\tau=10^{-2}$ to update caching decisions.
\end{itemize}

We evaluate the performance of all five algorithms with different values of edge server caching limit $Z$, service rate $\phi$, and installation cost parameter $\beta$. We choose $\phi=60$, $Z=6$, and $\beta=100$ if they are not specified. In addition, we present the regret of all four online algorithms in each time slot.

\subsection{Evaluation Results}

\begin{figure*}[t]
    \captionsetup[subfigure]{labelformat=empty,justification=centering,farskip=2pt,captionskip=1pt}
    \centering
    \begin{tabular}{cc}
        \subfloat[(a) Cost per time slot vs. service rate]
        {
           \includegraphics[width=0.235\linewidth]{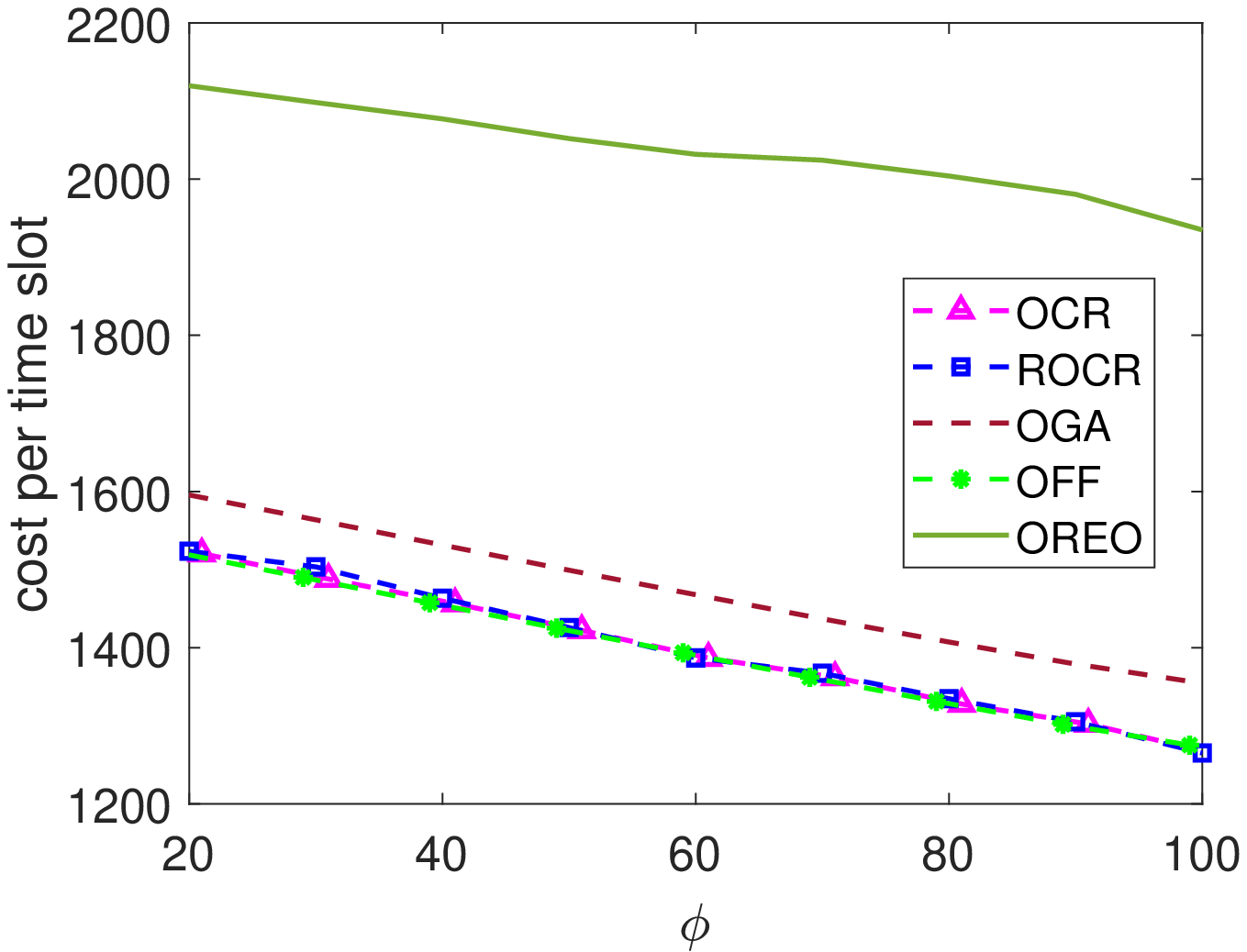}
           \label{fig:syn_phi}
        }
        \subfloat[(b) Cost per time slot vs. cache size]
        {
           \includegraphics[width=0.235\linewidth]{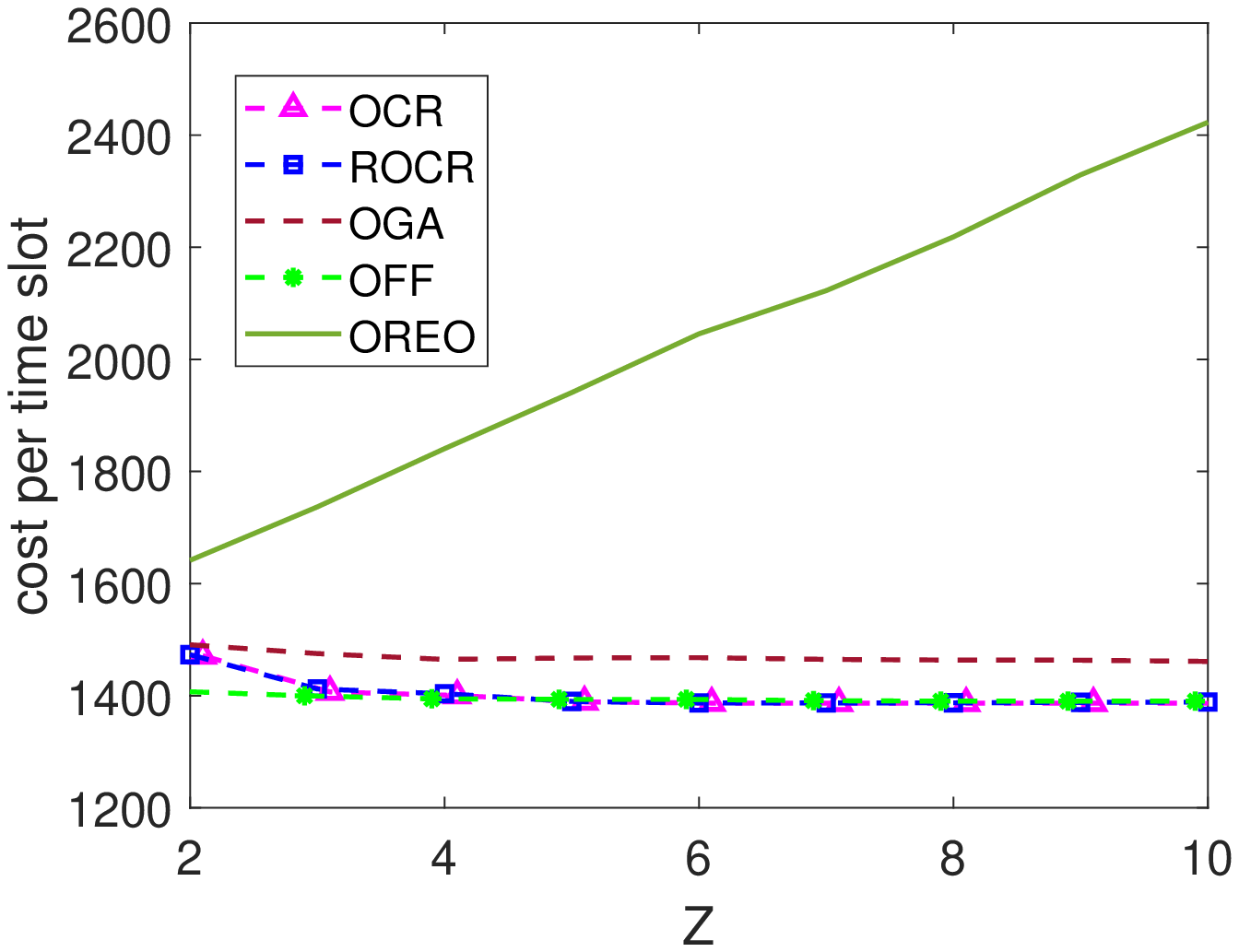}
           \label{fig:syn_z}
        }
        \subfloat[(c) Cost per time slot vs. installation cost parameter]
        {
           \includegraphics[width=0.235\linewidth]{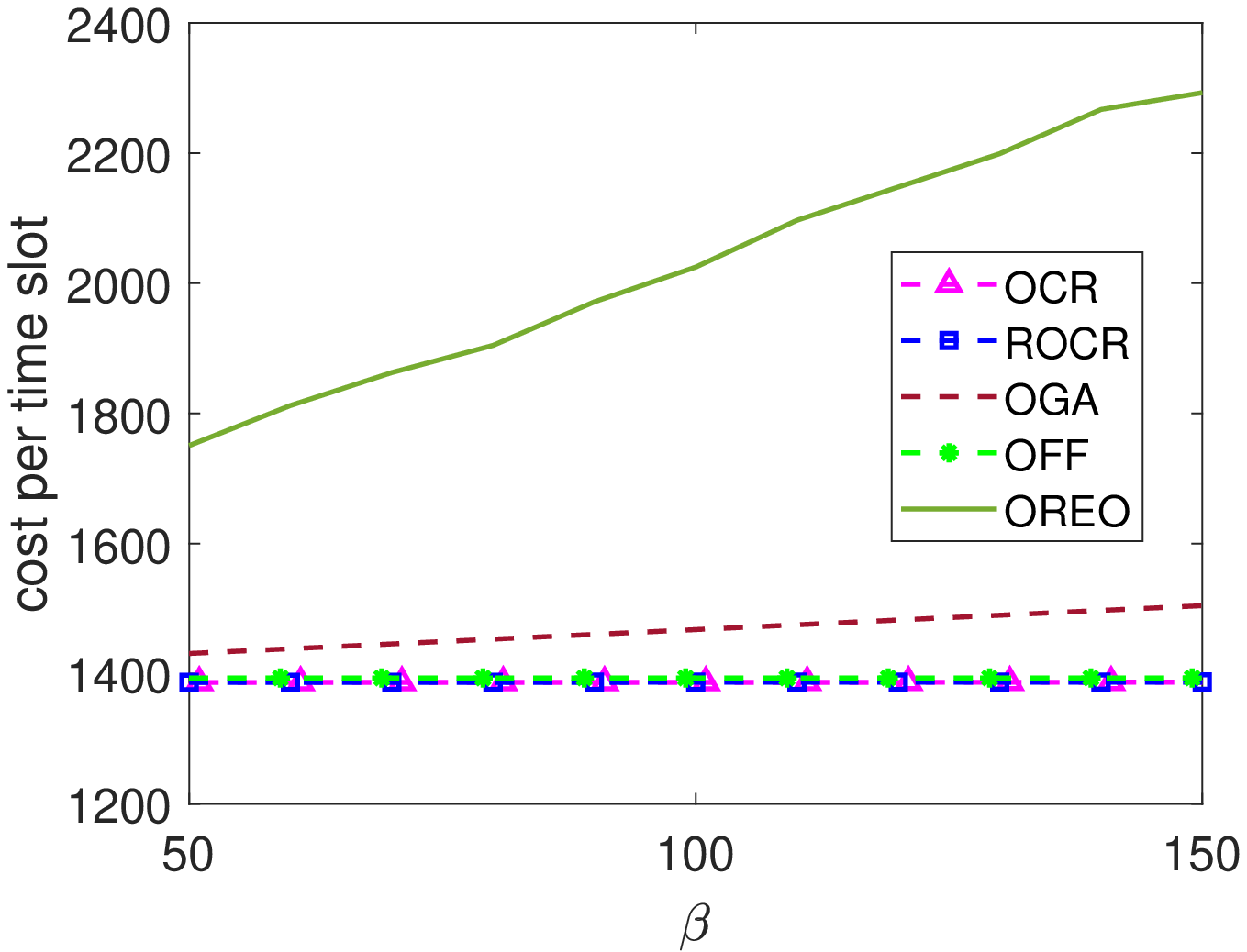}
           \label{fig:syn_beta}
        }
        \subfloat[(d) Regret per time slot vs. total time slot]
        {
           \includegraphics[width=0.235\linewidth]{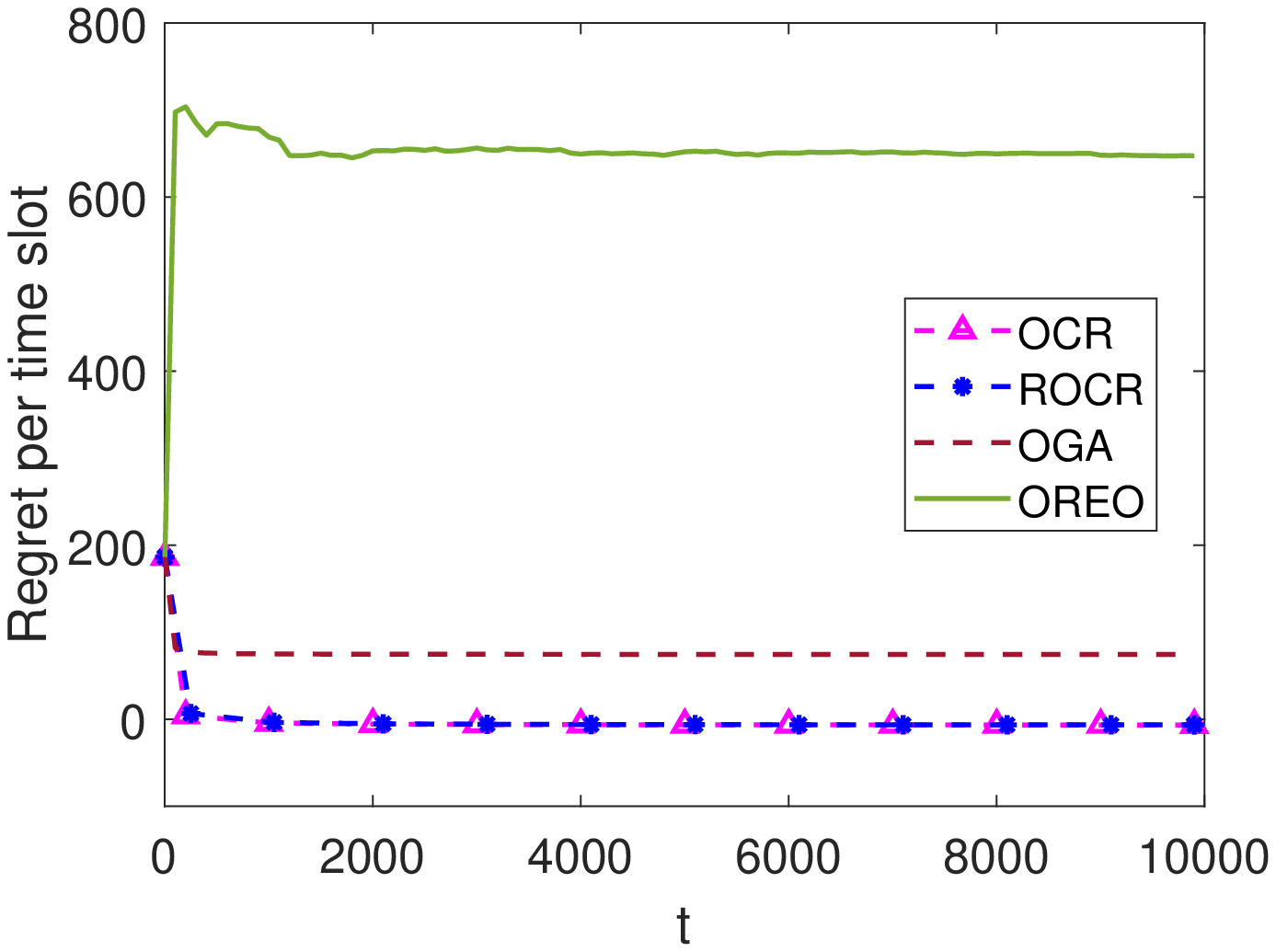}
           \label{fig:syn_regret}
        }
    \end{tabular}
    \vspace{-3mm}
    \caption{Simulation results using synthetic data.}
    \label{fig:sim:syn}
    \vspace{-1.5em}
\end{figure*}

\begin{figure*}[t]
    \captionsetup[subfigure]{labelformat=empty,justification=centering,farskip=2pt,captionskip=1pt}
    \centering
    \begin{tabular}{cc}
        \subfloat[(a) Cost per time slot vs. service rate]
        {
           \includegraphics[width=0.235\linewidth]{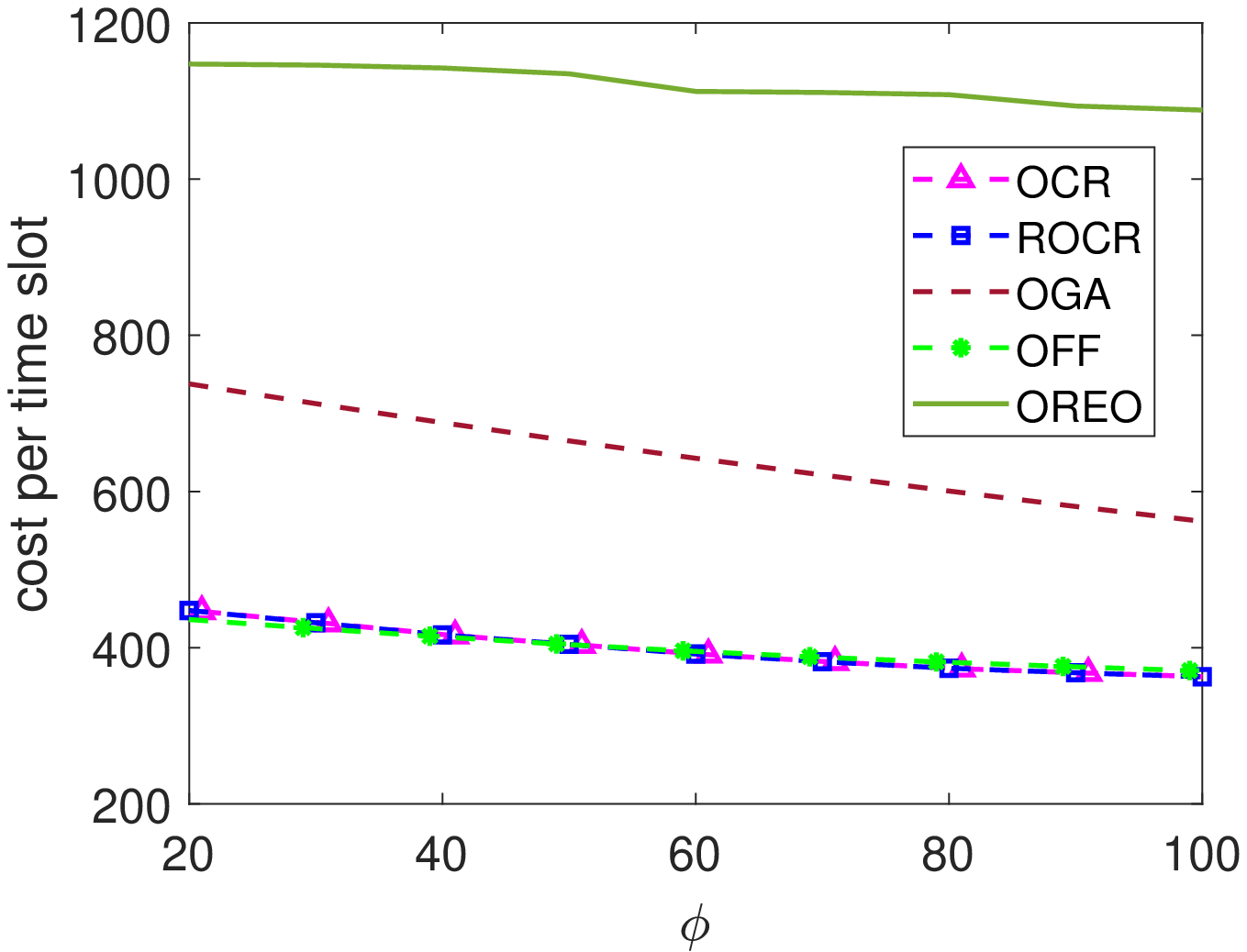}
           \label{fig:google_phi}
        }
        \subfloat[(b) Cost per time slot vs. cache size]
        {
           \includegraphics[width=0.235\linewidth]{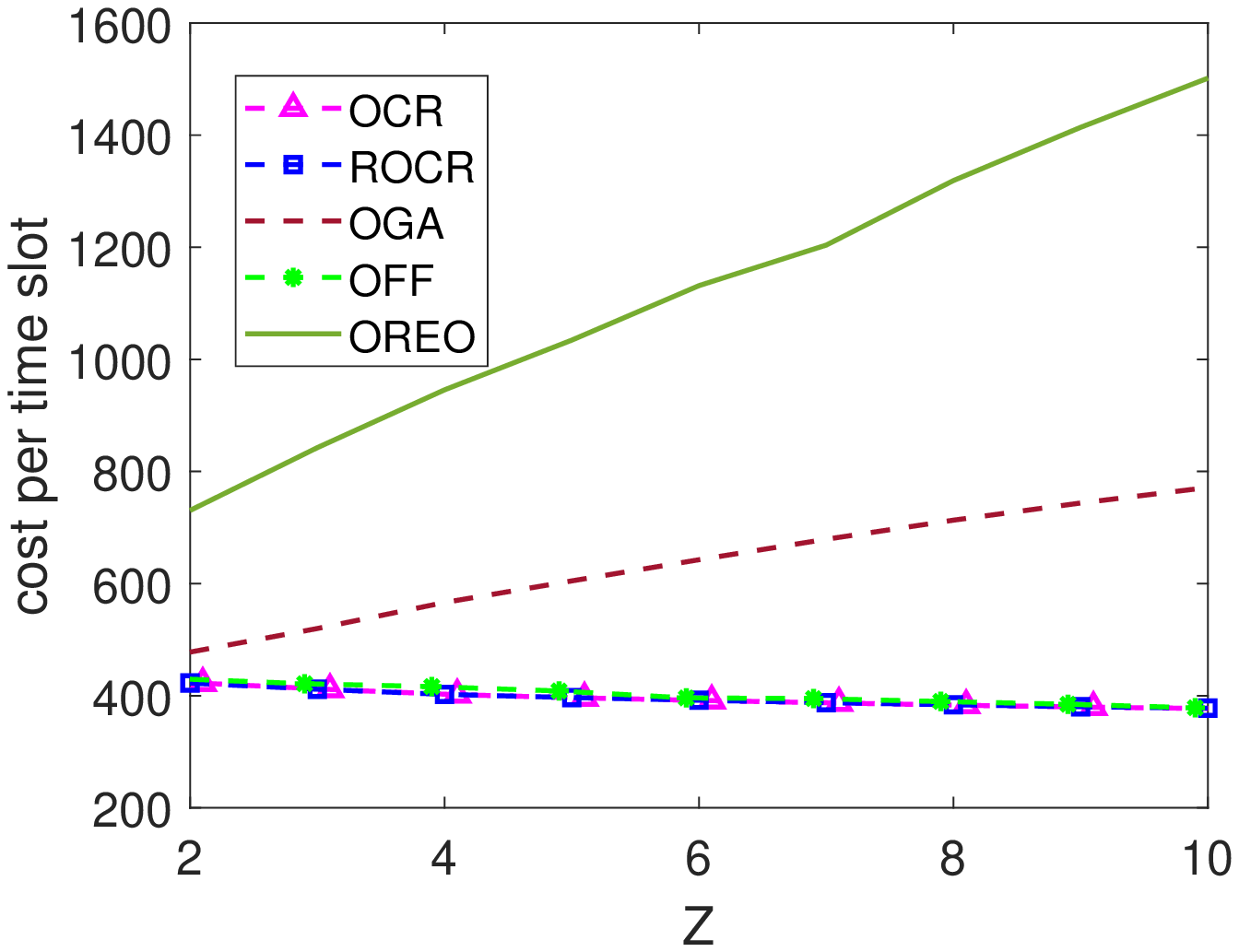}
           \label{fig:google_z}
        }
        \subfloat[(c) Cost per time slot vs. installation cost parameter]
        {
           \includegraphics[width=0.235\linewidth]{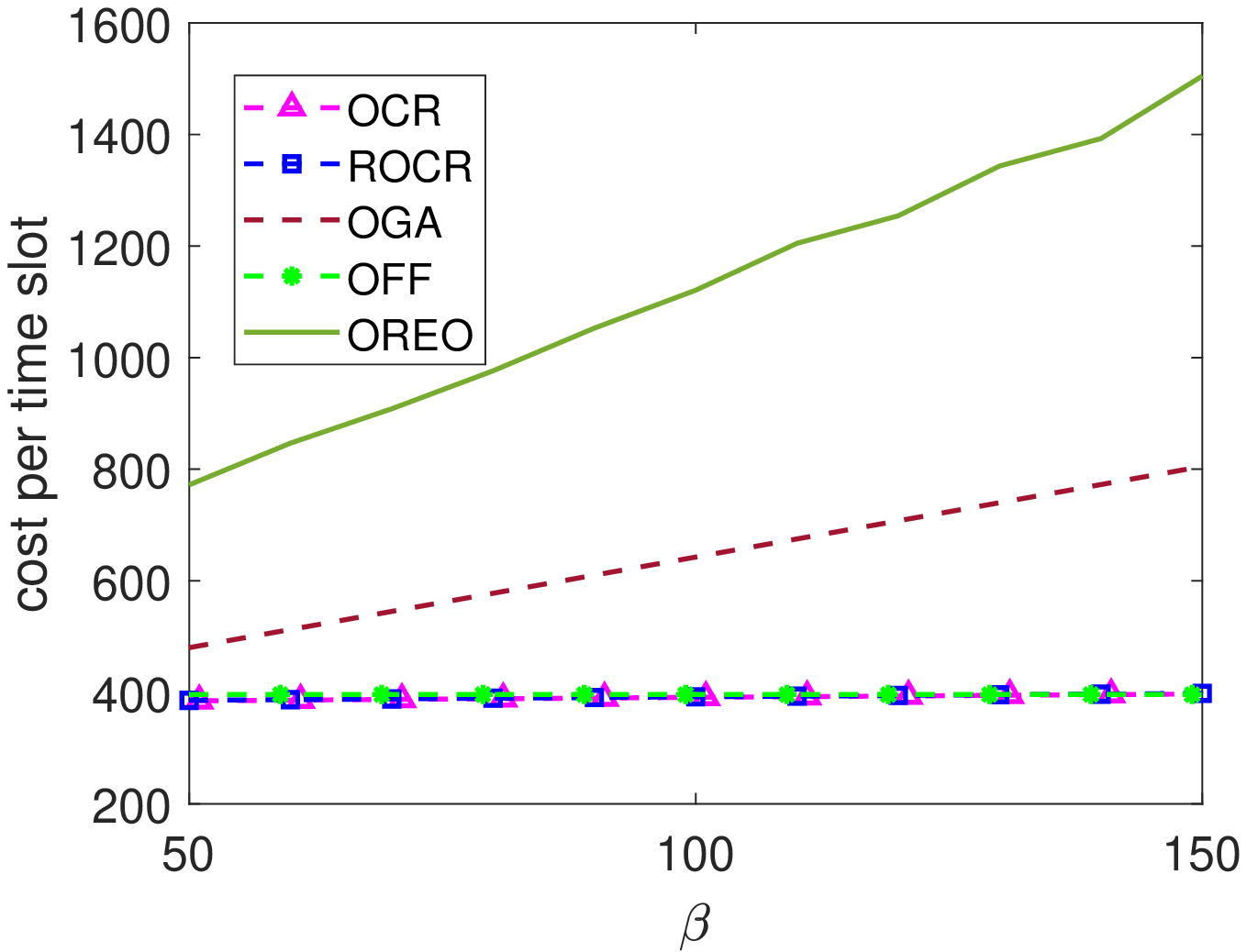}
           \label{fig:google_beta}
        }
        \subfloat[(d) Regret per time slot vs. total time slot]
        {
           \includegraphics[width=0.235\linewidth]{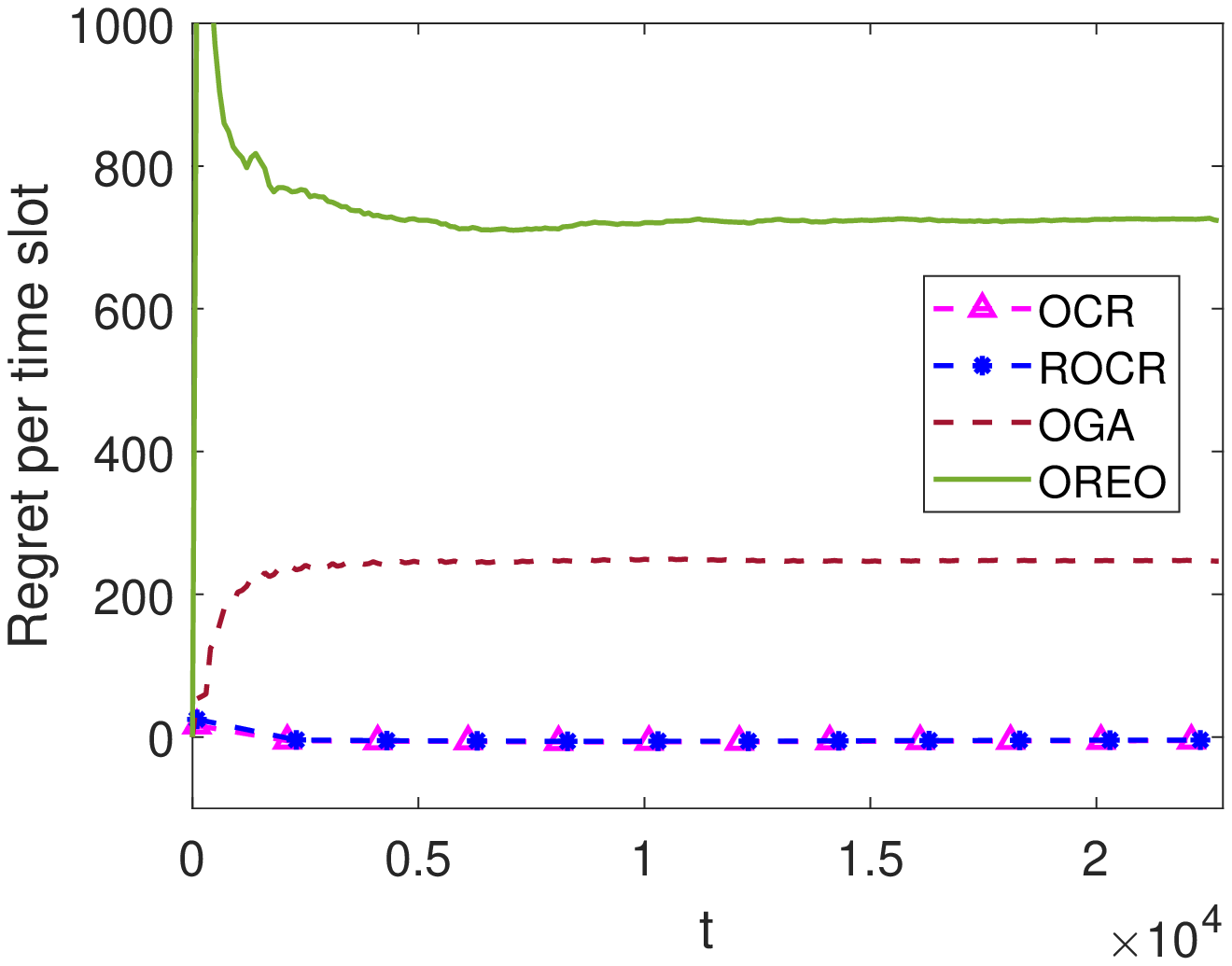}
           \label{fig:google_regret}
        }
    \end{tabular}
    \vspace{-3mm}
    \caption{Simulation results using the Google trace data.}
    \label{fig:simulation:google_heter}
    \vspace{-5mm}
\end{figure*}

The simulation results for two scenarios are shown in Fig.~\ref{fig:sim:syn} and Fig.~\ref{fig:simulation:google_heter}, and we can obtain several important observations. 

First, our ROCR significantly outperforms OREO in all settings. Though ROCR and OREO all jointly optimize service caching and routing, OREO assumes request arrival patterns are predictable and uses Gibbs sampling for cache updates, which causes massive installation cost and the surprising cost increment when we increase the cache size Z. Second, ROCR also outperforms OGA in all scenarios. While both algorithms are based on online gradient methods, ROCR can achieve better performance because it explicitly considers the processing latency and avoids the redundant cache changes when the edge server cannot process requests for popular services locally. Observations above show that any online algorithm for edge computing needs to address both memory and computation power constraints of edge servers as well as the challenge of unknown future requests. 

Third, our OCR has virtually the same performance as the optimal static offline policy and achieves nearly zero regrets, which is consistent with our analysis.

Finally, we note that ROCR and OCR have very similar performances in all cases. OCR produces fractional solutions for the service caching problem, and then ROCR transforms such fractional solutions into randomized solutions with integer solutions on every sample path. As discussed in Section~\ref{section:integral}, by carefully choosing which services to host at the edge on every sample path, ROCR is able to incur an installation cost that is at most three times larger than that of OCR. Our simulation results further show that the overall costs of ROCR and OCR are almost identical in practical scenarios.

\section{Conclusion}
\label{section:conclusion}

This paper studies the problem of service caching and routing without any knowledge about future requests. Motivated by a practical timescale separation, we formulate this problem as a two-stage online optimization problem that jointly considers the storage and computation constraints of the edge server, as well as the installation cost. We propose a low-complexity online algorithm for this problem that achieves sublinear regret bounds under a fractional relaxation. We further introduce a randomized algorithm that is guaranteed to produce integer solutions with provably small installation cost. Simulation results show that our ROCR and OCR algorithms have better performance than other recent proposed policies and achieve a similar performance as the optimal static offline policy.

\section*{Acknowledgment}

This material is based upon work supported in part by NSF under Award Number ECCS-2127721, in part by the U.S. Army Research Laboratory and the U.S. Army Research Office under Grant Number W911NF-18-1-0331, and in part by Office of Naval Research under Contract N00014-21-1-2385.

\bibliographystyle{ieeetr}
\bibliography{reference}

\end{document}